\documentclass[aps,prd,superscriptaddress,11pt,showpacs,notitlepage]{revtex4}
	\usepackage{graphicx}
	\usepackage{amsmath,amssymb,mathrsfs}
\usepackage{epsf}
\usepackage{epsfig}
\usepackage{subfig}

\usepackage{amsmath}

\usepackage{amssymb}

\numberwithin{equation}{section}

\newcommand{\be}{\begin{equation}}
\newcommand{\ee}{\end{equation}}
\newcommand{\bea}{\begin{eqnarray}}
\newcommand{\eea}{\end{eqnarray}}

\newcommand{\bb}{\bibitem}
 
\newcommand{\eqn}{\begin{eqnarray}}
\newcommand{\eqnx}{\end{eqnarray}}

\begin{document}
\title{Adding crust to BPS Skyrme neutron stars}
\author{C. Adam}
\affiliation{Departamento de F\'isica de Part\'iculas, Universidad de Santiago de Compostela and Instituto Galego de F\'isica de Altas Enerxias (IGFAE) E-15782 Santiago de Compostela, Spain}
\author{J. Sanchez-Guillen}
\affiliation{Departamento de F\'isica de Part\'iculas, Universidad de Santiago de Compostela and Instituto Galego de F\'isica de Altas Enerxias (IGFAE) E-15782 Santiago de Compostela, Spain}
\author{R. Vazquez}
\affiliation{Departamento de F\'isica de Part\'iculas, Universidad de Santiago de Compostela and Instituto Galego de F\'isica de Altas Enerxias (IGFAE) E-15782 Santiago de Compostela, Spain}
\author{A. Wereszczynski}
\affiliation{Institute of Physics,  Jagiellonian University,
Lojasiewicza 11, Krak\'{o}w, Poland}

\begin{abstract}
The Skyrme model and its generalisations provide a conceptually appealing field-theory basis for the description of nuclear matter and, after its coupling to gravity, also of neutron stars. 
In particular, a specific Skyrme submodel,   the so-called 
Bogomol'nyi-Prasad-Sommerfield  (BPS) Skyrme model,
allows both for an exact field-theoretic and a mean-field treatment of neutron stars, as a consequence of its perfect fluid property. A pure BPS Skyrme model description of neutron stars, however, only describes the neutron star core, by construction. Here we consider different possibilities to extrapolate a BPS Skyrme neutron star at high baryon density to a description valid at lower densities. In the exact field-theoretic case, a simple effective description of the neutron star crust can be used, because the exact BPS Skyrme neutron star solutions formally extend to sufficiently low densities.
In the mean-field case, on the other hand, the BPS Skyrme neutron star solutions always remain above the nuclear saturation density and, therefore, must be joined to a different nuclear physics equation of state already for the outer core. We study the resulting neutron stars in both cases,
 facilitating an even more complete comparison between Skyrmionic neutron stars and neutron stars obtained from other approaches, as well as with observations.
\end{abstract}
\maketitle 
\section{Introduction}
The Skyrme model \cite{skyrme} is a non-linear field theory of pion fields which represents a particular proposal for a low-energy effective field theory (EFT) of strong-interaction physics. Baryons are realized as topological solitons ("skyrmions") supported by the model \cite{manton-book}-\cite{rho-zah-book}.    
Already the original Skyrme model incorporates several nontrivial features of QCD in its defining properties, like chiral symmetry and its breaking, the conservation of baryon charge, or the extended character of nucleons. Further, it leads to a reasonable description of the physical properties of nucleons \cite{anw} and some light nuclei \cite{braaten}-\cite{light}. Several features, however, impede its use as a  quantitatively precise EFT of nuclear and strong-interaction physics. First of all, it leads to too high (classical) binding energies of nuclei. Secondly, some skyrmions are more symmetric than the nuclei they are supposed to descibe and do not reproduce the alpha-particle clustering observed in physical nuclei. Related to this issue, also the description of nuclear excitation spectra is only partially successful. 

The Skyrme model permits, in principle, generalizations via the inclusion of higher order terms or by the incorporation of further mesons into the theory, and both of these generalizations were considered \cite{nappi1}, \cite{castill}, \cite{weigel1} almost immediately after the interest in the model revived owing to the influential papers \cite{thooft}.
In the last years, moreover, it was found that among these generalizations of the Skyrme model particular cases can be identified \cite{BPS}-\cite{SutNaya1} which significantly improve on the shortcomings of the original model. In addition, new methods have been employed for their physical applications \cite{carbon}-\cite{bja-hal} which further improve this situation. On the one hand, several generalizations of the Skyrme model have been proposed which alleviate the binding energy problem. As demonstrated very recently, the particular coupling to vector mesons proposed in \cite{SutBPS}, in addition, also leads to the desired alpha particle cluster structures \cite{SutNaya2}. On the other hand, a more refined treatment of quantum excitations of skyrmions, beyond the rigid rotor approximation, has led to a vastly improved description of some nuclear excitation spectra \cite{halcrow1}-\cite{bja-hal}.  

In line with its role as an EFT for nuclear matter, after its coupling to gravity the Skyrme model should be capable of describing neutron stars (NS). The simplest ("hedgehog") ansatz, however, leads to stable solutions only for baryon number one \cite{kli}, \cite{bizon},  like in the case without gravity.
Further attempts to describe NS used variants of the rational map approximation \cite{piette1}, \cite{piette2} or a cubic lattice of alpha particles \cite{piette3} (the ground state of the standard Skyrme model for large baryon number \cite{footnote1}, \cite{kugler}, \cite{crystal}, \cite{vento}, \cite{half-sk}). 
This last attempt, in particular, already provided a reasonably good description like, e.g., a maximal NS mass of about 1.9 solar masses (for a recent review of skyrmionic NS we refer to \cite{naya}). 

In view of the improvements achieved by generalizations of the Skyrme model, their use for NS seems to be an obvious next step. If we restrict the field content to pions and demand a Lorentz-invariant lagrangian with a standard hamiltonian (quadratic in time derivatives), then the most general Skyrme model is
\be \label{gen-sk}
\mathcal{L}=\mathcal{L}_2+\mathcal{L}_4+\mathcal{L}_6 +\mathcal{L}_0
\ee
where the first two terms represent the sigma model (kinetic) term and the Skyrme term,
\be
\mathcal{L}_2= \lambda_2 \mbox{Tr} \; \partial_\mu U \partial U^\dagger, \;\; \mathcal{L}_4=\lambda_4 \mbox{Tr} ([L_\mu, L_\nu])^2 .
\ee
Here, $U$ is an $SU(2)$ valued matrix Skyrme field and $L_\mu = U^\dagger \partial_\mu U$ is the left-invariant Maurer-Cartan current. Further, the $\lambda_n$ are non-negative, dimensionful coupling constants. Moreover,  $\mathcal{L}_0=-\lambda_0 \mathcal{U} (\mbox{Tr} \, U)$ is a potential. The dependence on $\mbox{Tr} \, U$ only implies that isospin remains a symmetry, while chiral symmetry is broken. One frequent choice is the pion mass potential $\mathcal{U}=\mathcal{U}_\pi = (1/2) \mbox{Tr} \,(\mathbb{I}- U)$.  
The last term
\be
\mathcal{L}_6= -(24\pi^2)^2 \lambda_6 \mathcal{B}_\mu \mathcal{B}^\mu 
\ee
is just the baryon current squared,
\be \label{top-curr}
\mathcal{B}^\mu = \frac{1}{24\pi^2} \epsilon^{\mu \nu \rho \sigma} \mbox{Tr} \; L_\nu L_\rho L_\sigma, \;\;\; B= \int d^3 x \mathcal{B}^0 
\ee
where the topological charge $B$ is identified with the baryon number. For the Skyrme field $U$, we frequently use the parametrization $U(x)=e^{i \xi (x) \vec{n}(x) \cdot \vec{\tau}} $ where $\xi$ is called the profile function, $n^a$ is a unit isovector, and $\tau^a$ are the Pauli matrices. Further, we define
$h=(1/2)(1-\cos \xi )$ for convenience.

The description of NS using the full generalized model (\ref{gen-sk}) is, however, difficult. A full field-theoretic treatment of the self-gravitating generalized Skyrme model is beyond current possibilities for the large baryon numbers involved. Even the ground state of the model (\ref{gen-sk}) for large baryon number, as well as possible phase transitions at different densities, are currently unknown. There exists, however, a specific submodel \cite{BPS} of (\ref{gen-sk}) which leads to a drastic simplification and, at the same time, already to a rather realistic description of NS \cite{BPS-NS1}, \cite{BPS-NS2} (for an overview see \cite{naya}, \cite{rho-book}). The resulting NS are still compatible with the most important observational constraints. This so-called BPS (Bogomol'nyi-Prasad-Sommerfield) Skyrme model $\mathcal{L}_{\rm BPS} \equiv \mathcal{L}_6+\mathcal{L}_0$ has several features which make it an interesting model for certain bulk properties of nuclear matter and, in particular, of neutron stars. First of all, it is a perfect fluid already at the microscopic (field theoretical) level, without any need for a thermodynamical or hydrodynamical limit. A thermodynamical (mean-field) limit may, nevertheless, be performed easily \cite{term}, \cite{rho-book}. Secondly, a topological bound (BPS bound) for the static energy can be derived, and there exist infinitely many BPS solutions saturating the bound, with energies proportional to the topological charge \cite{BPS}, \cite{rho-book}.  The resulting classical nuclear binding energies are, therefore, zero, and small, realistic binding energies can be achieved by including further small corrections to the energy (spin, isospin, Coulomb energy, \ldots ) \cite{nearBPS}, \cite{rho-book}. Thirdly, the sextic term $\mathcal{L}_6$ provides the leading contribution to the energy and equation of state (EoS) for the generalized Skyrme model (\ref{gen-sk})  in the limit of large density, and this contribution to the EoS exactly coincides with the (leading) contribution induced by the omega meson repulsion in relativistic mean field theories (RMF) of nuclear physics \cite{Sk-eos}, like the Walecka model \cite{walecka}, \cite{schmitt}. This fact, together with the perfect-fluid property of the BPS Skyrme model, is the underlying reason for its success in the description of the central regions of NS, which provide the main contributions to their bulk properties. We want to emphasize that this possibility to describe the central, high-density region of NS by a simple and well-motivated physical model is particularly important, because the properties of baryonic matter at these high densities are still poorly understood, whereas low density regions can be described by standard methods of nuclear physics. The relevance of skyrmionic NS is further underlined by the observation that certain generic consequences of the Skyrme model, like a rather stiff EoS at high density or relatively large maximum masses $M_{\rm max} >2M_\odot$ of NS, are supported by recent observations, disfavoring models with too soft EoS.

The arguments given above already indicate that
a more complete and more detailed description of NS, including their peripheral, low density regions,  probably cannot be achieved by the BPS Skyrme model alone and needs a completion in terms of standard nuclear physics. Indeed,
\begin{itemize}
\item[{\em i)}]
Skyrmions for arbitrary baryon number are either of a strictly finite extension ("compactons", like in the case of the BPS Skyrme model for a wide class of potentials), or have exponential tails. In both cases they essentially describe finite chunks of nuclear matter, already in the absence of gravity.
\item[{\em ii)}] Classical soliton solutions (skyrmions) have a given baryon number but do not distinguish between protons and neutrons (after quantization, the proton and neutron content of a skyrmion is determined by its isospin).
\item[{\em iii)}] Classical BPS skyrmions solve a first-order BPS equation and, therefore, have identically zero pressure everywhere. 
\item[{\em iv)}] Classical BPS skyrmions, therefore, should be interpreted as describing symmetric nuclear matter.  The BPS property also excludes surface effects (the energy of a BPS skyrmion is exactly proportional to the volume). If electromagnetic effects are not taken into account, either (which we assume in this paper), then BPS skyrmions of sufficiently large baryon number describe infinite nuclear matter at saturation.
\item[{\em v)}] This implies that, as long as we model nuclear matter in terms of the BPS Skyrme model only (without the inclusion of further terms), the model parameters should be calibrated to infinite nuclear matter. That is to say, its soliton solutions should reproduce the nuclear saturation density $n_0 = 0.160\, \mbox{fm}^{-3}$ and the energy per nucleon of infinite nuclear matter $E_{\rm inm} = m_{\rm N} - E_{\rm b} = (938.9 - 16.0)\,  \mbox{MeV} = 922.9\, \mbox{MeV}$ (here $m_{\rm N}$ is the nucleon mass and $E_{\rm b}$ is the binding energy per nucleon of infinite nuclear matter; we use the up-to-date values given in \cite{vinas}, table 2). The average baryon density (baryon density in mean-field theory) is, thus, equal to $n_0$ for BPS skyrmions.
\end{itemize} 
As a consequence of the above, the baryon density of BPS Skyrme NS within a mean-field theory (MFT) approach is bounded from below by $n_0$ and takes the value $n_0$ at the NS surface. Physically, describing a NS purely by the BPS Skyrme model within MFT
implies that only the infinite (or symmetric) nuclear matter aspects of NS matter are modeled. For $n>n_0$, we still have different equations of state (EoS) for different MFT BPS Skyrme models, which are related to different choices for the potential $\mathcal{U}$, see section IV.   In principle, already in the region of high density $n>n_0$, isospin corrections to the BPS Skyrme NS masses should be considered, to account for the mainly neutron nature of nucleons in a NS. Numerical calculations, however, indicate that these isospin corrections are small.  More importantly, it is well-known from standard nuclear physics calculations that NS contain peripheral regions of lower density $n<n_0$. This implies that the equation of state (EoS) resulting from the MFT BPS Skyrme model must be joined to a standard nuclear physics EoS at some point $n_* >n_0$. The NS core is, thus, divided into an inner core $n>n_*$, described by the MFT BPS Skyrme model and an outer core described by a standard nuclear physics EoS. Concretely, we shall use the universal EoS of \cite{vinas}, which is based on a Brueckner-Hartree-Fock many-body calculation for the core and the BCPM (=Barcelona-Catania-Paris-Madrid) nuclear energy density functional for the crust (BCPM EoS for short).

The situation is slightly different for the exact field theory solutions of BPS Skyrme NS. First of all, the perfect fluid described by the exact BPS Skyrme model (beyond MFT) is {\em non-barotropic}, and an algebraic EoS relating the energy density $\rho$ and the pressure $p$ does not exist, see section IV.
A low-density completion of an exact BPS Skyrme NS, therefore, cannot be achieved by simply joining different EoS. Secondly, the microscopic baryon density of an exact BPS Skyrme NS is zero at the NS surface and, therefore, takes arbitrarily small values close to it. Still, for low densities a NS description in terms of the exact BPS Skyrme model will probably not be reliable, and a low-density completion is required. In particular, the BPS Skyrme model leads to a homogeneous matter distribution, whereas matter in the crust of a NS is known not to be homogeneous, essentially consisting of droplets of nuclear matter embedded in a gas of nucleons and electrons. A rather obvious proposal, thus, consists in using the exact BPS Skyrme NS for the core region and a different description for the crust region. The core-crust (cc) transition typically occurs for baryon densities $n_{\rm cc} \sim (1/2) n_0$, where the precise value of $n_{\rm cc}$ is slightly model dependent. The above proposal, however, meets two obstacles, namely the difficulty in joining an exact BPS Skyrme NS with a NS derived from an EoS mentioned above, and the intrinsic difficulty of a full microscopic description of the inhomogeneous crust. To overcome these problems, we propose to use the effective description of the NS crust recently developed in \cite{ZFH}. This effective description cannot lead to a complete description of the complicated crust structure, but it reproduces the bulk observables of NS (masses, radii) with a surprisingly high precision. It is one of the aims of this paper to apply the effective crust description of \cite{ZFH} to the NS resulting from the exact BPS Skyrme model.

We briefly review the theoretical description of NS in the next section. In section III, we summarize the effective crust description of \cite{ZFH}. Some relevant properties of the BPS Skyrme model are introduced in section IV.   In section V, we present the numerical results 
for exact BPS Skyrme NS with the effective crust of  \cite{ZFH}, and 
for BPS Skyrme NS in MFT joined to the EoS of \cite{vinas}.
Finally, section VI contains our conclusions. 
We use units such that the speed of light $c=1$. We are, thus, left with a mass (or energy) unit and a length unit where, depending on the context, we use either nuclear physics units (MeV and fm) or stellar astrophysics units (solar masses $M_{\odot}$ and km). Finally, Newton's constant is 
$G_{\rm N}= 1.322 \cdot 10^{-42} \, {\rm fm} \, {\rm MeV}^{-1}$.
\section{Neutron Stars}
The theoretical description of a neutron star (NS) usually starts with the assumption that the matter composing it can be described by the energy-momentum tensor (EMT) of a perfect fluid,
\be \label{fluid-emt}
T^{\mu\nu} =  (p+\rho )u^\mu u^\nu - pg^{\mu\nu}.
\ee
Here, $u^\rho$ is the four-velocity of the fluid, $\rho $ is its energy density, and $p$ its pressure.  
For the purposes of the present paper, we are only interested in static NS solutions. In the corresponding static space-time, the time direction can always be chosen perpendicular to space-like hypersurfaces, implying a block-diagonal metric 
\be \label{bloc-metric}
ds^2 = g_{00} (\vec x)dt^2 - g_{ij} (\vec x)dx^i dx^j
\ee 
and the fluid four-velocity $u^\mu = (g_{00})^{-1/2} \delta^{0\mu}$.  
In such a static space-time, the Einstein equations $G_{\rho\sigma} = 8\pi G_{\rm N} T_{\rho \sigma}$ are compatible with the assumption of spherical symmetry, i.e., with a metric
\be \label{sym-metric}
ds^2 = {\bf A}(r) dt^2 - {\bf B}(r) dr^2 - r^2 (d\theta^2 + \sin^2 \theta d\varphi^2 )
\ee
and with a pressure $p(r)$ and energy density $\rho(r)$ which only depend on the radial coordinate $r$.
For this ansatz, the Einstein equations simplify to three independent ordinary differential equations (ODEs) for the four functions $\rho (r)$, $p(r)$, ${\bf A}(r)$ and ${\bf B}(r)$, known as Tolman-Oppenheimer-Volkoff (TOV) equations \cite{OV}, \cite{Tol}. It turns out that the function ${\bf A}$ can be completely eliminated from two of the three TOV equations, such that the system simplifies to a system of two equations for $\rho$, $p$ and ${\bf B}$, and a third equation which expresses ${\bf A}$ in terms of the remaining functions. Concretely, the first two equations can be expressed like (here $p' \equiv (dp/dr) $ etc.)
\bea \label{Tov1}
m' &=& 4\pi r^2 \rho , \qquad   {\bf B}(r) \equiv \left(1- \frac{2G_{\rm N}m(r)}{r} \right)^{-1} \\
\label{Tov2a}
p' &=&- \frac{\rho + p}{r} \left( \frac{1}{2} ({\bf B}-1 ) + 4\pi G_{\rm N} r^2 {\bf B}p\right) 
\\ \label{Tov2b}
&=& - \frac{G_{\rm N}(\rho + p)}{1 - \frac{2G_{\rm N}m(r)}{ r}}\left( \frac{ m(r)}{ r^2}+ 4\pi rp \right)
\eea
whereas the third equation is
\be \label{Tov3}
\frac{{\bf A}'}{{\bf A}} = \frac{1}{r}({\bf B}-1) + 8\pi G_{\rm N} r{\bf B}p. 
\ee
The covariant conservation of the EMT, $g_{\lambda\nu}\nabla_\mu T^{\mu\nu}=0$, which 
for a perfect fluid reads
\be \label{cov-cons}
\frac{\partial_\lambda p}{\rho + p} = -\frac{1}{2}\partial_\lambda \ln \sqrt{g_{00}},
\ee
is, in fact, implied by the above Einstein equations. Indeed, Eqs. (\ref{Tov2a}) and (\ref{Tov3}) immediately lead to
\be \label{TOV3a}
\frac{p'}{\rho + p} = -\frac{1}{2} \frac{{\bf A}'}{{\bf A}}
\ee
which is just Eq. (\ref{cov-cons}) for our spherically symmetric ansatz.   Eqs. (\ref{Tov1}) and (\ref{Tov2a}) are two equations for the three unknown functions $\rho$, $p$ and ${\bf B}$, therefore a third equation is required to close the system.
The simplest possibility, which is assumed in almost all investigations of NS, is to consider an algebraic equation of state (EoS) $\rho = \rho (p)$. This is equivalent to the assumption that the fluid described by (\ref{fluid-emt}) is barotropic. Another possibility is that the field theory describing the NS matter is already of the perfect-fluid form, and then the system of equations is closed by the corresponding field equations. If, in addition, these field equations are equivalent to the covariant energy-momentum conservation condition (\ref{cov-cons}), then they are implied by the Einstein equations and the system (\ref{Tov1}) and (\ref{Tov2a}) closes by itself.
This happens, e.g., for a field theory of one real scalar field, or if the field space is effectively one-dimensional after a symmetry reduction to spherical symmetry. Most field theories are not of the perfect-fluid form and require a macroscopic description like mean-field theory (MFT) to arrive at a perfect-fluid energy momentum tensor and an EoS. The BPS Skyrme model, on the other hand, is a perfect fluid and, therefore, offers the unique opportunity to compare exact and MFT results. 

\section{Effective Crust Description}
\subsection{Baryon chemical potential}
The baryon number chemical potential $\boldsymbol{\mu}$ is one important observable for the description of NS. In addition, it plays a distinguished role in the effective crust description of \cite{ZFH}, therefore it will be useful to briefly review some of its properties. There exist several definitions of the chemical potential, which are all equivalent in the case of a barotropic fluid where the energy density can be expressed as a function of the pressure, $\rho = \rho (p)$ or, equivalently, both $\rho$ and $p$ can be expressed as functions of the baryon number density $n$, i.e., $\rho = \rho (n)$, $p=p(n)$.   The baryon chemical potential is defined as the change of the free energy $F$ under a change of baryon number $B$ at constant volume $V$. For our purposes, we assume zero temperature $T=0$, such that the free energy coincides with the (static Skyrmion) energy, $E(V,B)=F(V,B,T=0)$. Consequently,
\be \label{mu-def1}
\boldsymbol{\mu} = \left. \frac{\partial E}{\partial B} \right|_V .
\ee
The pressure is spatially constant for matter in a thermodynamical equilibrium. For barotropic matter this implies that the energy density and baryon number density are constant, as well, and may be simply defined like $\rho = E/V$ and $n=B/V$. Eq. (\ref{mu-def1}) then leads to 
\be \label{mu-def2}
\boldsymbol{\mu} = \frac{d\rho}{dn} .
\ee
Further, if $\rho$ is considered as a function of $n$, then the pressure exerted by the matter with particle number density $n$ (in our case, baryons) is given by $p=n(d\rho/dn) -\rho$. Inserting definition (\ref{mu-def2}), we find yet another definition for $\boldsymbol{\mu}$,
\be \label{mu-def3}
\rho + p = n\boldsymbol{\mu}.
\ee
Finally, taking the differential of this expression, $dp =  \boldsymbol{\mu} dn + nd\boldsymbol{\mu} -d\rho = nd\boldsymbol{\mu}$ and replacing $n$ with the help of (\ref{mu-def3}), leads to the relation
\be \label{mu-def4}
\frac{dp}{p+\rho} = \frac{d\boldsymbol{\mu}}{\boldsymbol{\mu}}.
\ee
For a {\em non-barotropic} perfect fluid, the pressure in flat space is still constant in equilibrium, as a consequence of the conservation equation (\ref{cov-cons}). $\rho$ and $n$, on the other hand, do not have to be constant. As a result, the above expressions for $\boldsymbol{\mu}$ are no longer equivalent,  leading to the obvious question of which one to use to define the chemical potential. Some simple thermodynamical considerations allow us to answer this question. For a perfect fluid, the densities $\rho$, $n$ and $p$ are the natural variables, so definition (\ref{mu-def1}) does not apply directly. Further, without gravity (and in the absence of external potentials), both the temperature (zero in our case) and the chemical potential must be spatially constant in thermodynamical equilibrium. In particular, a non-constant chemical potential would allow to lower the energy by re-arranging the baryonic matter.
 But only definition (\ref{mu-def4}) implies a constant chemical potential for an equilibrium (constant pressure) configuration. Finally, in the presence of gravity, both the temperature \cite{tolman2} and the chemical potential \cite{klein} are  redshifted (no longer constant) in equilibrium, i.e., $T\sqrt{g_{00}}=$ const. and $\boldsymbol{\mu}\sqrt{g_{00}}=$ const. This may be re-expressed like $\partial_\lambda \ln \boldsymbol{\mu} = -(1/2)\partial_\lambda \ln \sqrt{g_{00}}$  and is, again, implied by the definition (\ref{mu-def4}) together with the conservation equation (\ref{cov-cons}). In particular, for spherical symmetry and using (\ref{TOV3a}), we get
 \be \label{mu'}
 \frac{\boldsymbol{\mu}'}{\boldsymbol{\mu}} = \frac{p'}{p+\rho} = - \frac{1}{2}\frac{\bf{A}'}{\bf{A}}.
 \ee
 We remark that Eq. (\ref{mu'}) implies that $\boldsymbol{\mu}$ is determined only up to a multiplicative constant, which does not follow from the TOV equations and must be fixed by some physical considerations. 
\subsection{Crust description of Zdunik, Fortin and Haensel}
The effective crust description of \cite{ZFH} is motivated by the problem that nuclear matter in the crust of a NS is not homogeneous, so finding a realistic EoS for it is difficult. In particular, only a few unified EoS, i.e, derived from the same nuclear model for the NS crust and core, are available. On the other hand, if the crust and core EoS stem from different nuclear models (non-unified EoS), then the matching at the core-crust (cc) interface introduces uncertainties, especially for the NS radius.
The procedure of \cite{ZFH} does not require a knowledge of the crust EoS, at all. All that is required are some NS core properties derivable from the core description (core mass, core radius, the chemical potential at the core radius (or cc interface), $\boldsymbol{\mu}_{\rm cc}$), and the chemical potential
at the NS surface (a physical input value). Further, a comparison with the full calculations for some unified EoS shows that the simple crust description of \cite{ZFH} reproduces certain bulk observables (NS mass, radius) with a surprisingly high precision.  

The procedure of \cite{ZFH} starts from the second TOV equation (\ref{Tov2b}), with the following  additional assumptions. i) the core contribution to the NS mass is much bigger than the crust contribution. For the crust region $r_{\rm cc} \le r \le R$, the mass function $m(r)$ may, therefore, be replaced by the full NS mass $M_{\rm NS} = m(R)$ (here $R$ is the NS radius where $p(R)=0$). ii) In the last term at the r.h.s. of (\ref{Tov2b}), $rp$ may be neglected in comparison with $M_{\rm NS}/r^2$ (the pressure close to the surface is small). With these assumptions, Eq. (\ref{Tov2b}), simplifies to
\be
\frac{dp}{p+\rho} = -G_{\rm N} M_{\rm NS} \frac{dr}{r^2 (1-2G_{\rm N} M_{\rm NS}/r)} .
\ee
Here, the l.h.s. is just the defining relation for the chemical potential, (\ref{mu-def4}). Further, the r.h.s. is a given function of $r$, independent of any EoS. Integrating both sides from $r_{\rm cc}$ to $R$ (from $p_{\rm cc} = p(r_{\rm cc})$ to $p(R)=0$), we arrive at
\be \label{crust-radius}
\left( \frac{\boldsymbol{\mu}_{\rm cc}}{\boldsymbol{\mu}_0} \right)^2 = \frac{1-(2G_{\rm N}M_{\rm NS})/R}{1-(2G_{\rm N}M_{\rm NS})/r_{\rm cc}}
\ee
where $\boldsymbol{\mu}_0 = \boldsymbol{\mu}(R)$ is the chemical potential at the NS surface. This equation permits to determine the NS radius $R$ from $\boldsymbol{\mu}_0$, $\boldsymbol{\mu}_{\rm cc}$, $r_{\rm cc}$ and $M_{\rm NS}$. Here, $\boldsymbol{\mu}_0$ and $\boldsymbol{\mu}_{\rm cc}$ are determined from physical considerations. $\boldsymbol{\mu}_0$ is the energy per baryon of iron, $\boldsymbol{\mu}_0 \sim 930\,$MeV. Further, calculations for different unified EoS determine $\boldsymbol{\mu}_{\rm cc}$ to be about $\boldsymbol{\mu}_{\rm cc} \sim 955\,$MeV \cite{ZFH}. This is the value we shall use in this paper, implying $(\boldsymbol{\mu}_{\rm cc}/\boldsymbol{\mu}_0) = 1.027$. $r_{\rm cc}$ is now determined from the core EoS. For a barotropic fluid in the core, $\boldsymbol{\mu}(r)$ may be determined directly from the solution of the TOV equations using the algebraic relation (\ref{mu-def3}).
$r_{\rm cc}$ then follows from $\boldsymbol{\mu}(r_{\rm cc}) = \boldsymbol{\mu}_{\rm cc} = 955\,$MeV.
For a non-barotropic fluid in the core, only the differential relation (\ref{mu-def4}) or (\ref{mu'}) is available, and the full function $\boldsymbol{\mu}(r)$ in the core (including the multiplicative constant) can only be determined by imposing one additional physical condition.

Finally, the NS mass can be expressed as a sum of the core and crust mass contributions, $M_{\rm NS} = M_{\rm core} + M_{\rm crust}$. Here, $M_{\rm core} = m_{\rm core} (r_{\rm cc})$ is the value of the mass function
$m_{\rm core} (r)$ for the core EoS evaluated at $r=r_{\rm cc}$. Further, taking into account the smallness of $M_{\rm crust}$,  $M_{\rm crust}/M_{\rm core} <<1$, it may be determined from the TOV equation (\ref{Tov2b}) by using an even cruder approximation. Indeed, neglecting $p$ also in $(p+\rho)$, replacing $m(r)$ by a constant $M$ and using $\rho = m'/(4\pi r^2)$ we get \cite{ZFH} 
\be
\frac{dm}{dp} =- \frac{4\pi r^4}{G_{\rm N}M}\left( 1 - \frac{2G_{\rm N}M}{r}\right) .
\ee
In the interval $r_{\rm cc} \le r \le R$, the r.h.s. of this expression does not vary too much. In a last approximation, we therefore assume that it is constant (i.e., its variation is a sub-leading effect) and given by its value at $r_{\rm cc}$. The reason for this choice is that, by assumption, we know the core EoS and, therefore, all observables at $r_{\rm cc}$. The integration is then trivial and leads to
\be \label{crust-mass}
M_{\rm crust} = \int_{p_{\rm cc}}^0  \frac{dm}{dp} =
\frac{4\pi p_{\rm cc} r_{\rm cc}^4}{G_{\rm N}M_{\rm core}}\left( 1 - \frac{2G_{\rm N}M_{\rm core}}{r_{\rm cc}}\right)
\ee
where $p_{\rm cc} \equiv p_{\rm core} (r_{\rm cc})$.

\section{BPS Skyrme model}
For a general metric, the BPS Skyrme model reads (we introduce the new coupling constants $\lambda$ and $\mu$ for convenience)
\be \label{bps-act}
S_{\rm BPS} = \int d^4 x |g|^\frac{1}{2} \left( -\lambda^2 \pi^4 |g|^{-1} g_{\rho\sigma} {\cal B}^\rho {\cal B}^\sigma - \mu^2 {\cal U} \right) ,
\ee
leading to the perfect fluid EMT 
\be
T^{\rho\sigma} = -2|g|^{-\frac{1}{2}}\frac{\delta}{\delta g_{\rho\sigma}} S_{\rm BPS} = (p+\rho )u^\rho u^\sigma - pg^{\rho\sigma}
\ee
with
\bea
u^\rho &=& {\cal B}^\rho / \sqrt{g_{\sigma \pi} {\cal B}^\sigma {\cal B}^\pi} \\
\rho &=& \lambda^2 \pi^4 |g|^{-1} g_{\rho\sigma} {\cal B}^\rho {\cal B}^\sigma + \mu^2 {\cal U} \\
p &=& \lambda^2 \pi^4 |g|^{-1} g_{\rho\sigma} {\cal B}^\rho {\cal B}^\sigma - \mu^2 {\cal U} .
\eea
Here, we use the convention that the totally anti-symmetric $\epsilon$ symbol used in the definition (\ref{top-curr}) of the baryon density current $\mathcal{B}^\mu$ still obeys $\epsilon_{0123} = - \epsilon^{0123} =1$, even for a general metric. That is to say, the $\epsilon$ symbol and, as a consequence, also $\mathcal{B}^\mu$ are tensor densities rather than tensors, and $\mathcal{B}^\mu$ still obeys the ordinary conservation law $\partial_\mu \mathcal{B}^\mu =0$. This explains the additional factor  $|g|^{-1}$ in Eq. (\ref{bps-act}). The true tensor $\tilde{\mathcal{B}}^\mu$ (contravariant vector, obeying $\nabla_\mu \tilde{\mathcal{B}}^\mu =0$) is  $\tilde{\mathcal{B}}^\mu = (1/\sqrt{|g|}) \mathcal{B}^\mu$. 

As the BPS Skyrme model is a perfect fluid, it can be inserted directly into the TOV equations, without the need for a thermodynamical or hydrodynamical limit. Indeed, 
the ansatz
\be \label{axi-sym}
h = h (r), \quad \vec n = (\sin \theta \cos B\varphi , \sin \theta \sin B \varphi , \cos \theta )
\ee
for the Skyrme field leads to a spherically symmetric energy density (in the standard Skyrme model the same ansatz only preserves an axial symmetry of the energy density \cite{weigel2}). Further, for the BPS Skyrme model this ansatz together with the spherically symmetric metric  (\ref{sym-metric}) 
are is compatible with the Einstein and Skyrme field equations and lead to the $r$-dependent energy density and pressure expressions
\bea \label{full-rho}
 \rho  & =& \frac{4 B^2 \lambda^2}{{\bf B}r^4} h(1-h)h'^2 + \mu^2 {\cal U}(h) , \\ \label{full-p}
 p & =& \frac{4 B^2 \lambda^2}{{\bf B}r^4} h(1-h)h'^2 - \mu^2 {\cal U}(h) .
 \eea 
 The boundary conditions are $h(0)=1$ and $h( R)=0$ (or $\xi (0)=\pi$, $\xi (R)=0$), where $R$ is the radius where the skyrmion approaches its vacuum value (the NS surface).
Further, we used that for the above ansatz $\mathcal{B}^0 =- (B/2\pi^2) \sin \theta \sin^2 \xi \, \xi' $, where $\sin^2 \xi \, \xi' = 4h' \sqrt{h(1-h)}$ and $\xi' \equiv \partial_r \xi$, etc. For a block-diagonal metric (\ref{bloc-metric}) and for static fields we may define the thermodynamical baryon density $n$ via
\be
\mathcal{B}^0 = \sqrt{|g|}\tilde{\mathcal{B}}^0 = \sqrt{g^{(3)}} n \, , \quad B = \int d^3 x \sqrt{g^{(3)}} n
\ee
where $g^{(3)}$ is the determinant of the spatial metric $g_{ij}$. For the above ansatz, this leads to
\be
\mathcal{B}^0 = r^2\sqrt{\bf{AB}}\sin \theta \, \tilde{\mathcal{B}}^0 (r) = r^2\sqrt{\bf{B}}\sin \theta \, n(r)
\; , \quad n(r) = -\frac{2B}{\pi^2 r^2 \sqrt{\bf{B}}}h' \sqrt{h(1-h)}
\ee
and to the (off-shell) thermodynamical relation 
\be \label{rho-p-n-rel}
\rho + p = 2\pi^4 \lambda^2 n^2.
\ee
Despite this relation, the perfect fluid described by the BPS Skyrme model is {\em non-barotropic}. In particular, in flat (Minkowski) space-time equilibrium configurations (static solutions) lead to constant pressure but to a non-constant energy density and baryon density, see below.

\subsection{Thermodynamics of the BPS Skyrme model}
For the purpose of this section, we will only consider static Skyrme configurations $U(\vec x)$ in flat (Minkowski) space, where we shall, however, allow for arbitrary space coordinates (an arbitrary flat metric $g_{ij}$) on $\mathbb{R}^3$. Further, the field space (target space) SU(2), as a manifold, is just the unit three-sphere $\mathbb{S}^3$, therefore each static Skyrme configuration defines a map $U: 
\mathbb{R}^3 \to \mathbb{S}^3$. The baryon density allows us to define the following three-form,
\be \label{three-form}
\mathcal{B}^0 d^3 x = n\, \sqrt{g^{(3)}} d^3 x \equiv n\,   {\rm vol}_{\mathbb{R}^3}
\ee
where ${\rm vol}_{\mathbb{R}^3}$ is the invariant volume form on ${\mathbb{R}^3}$. Here, the domain of the three-form (\ref{three-form}) is restricted to the region where $n\not= 0$. The topological character of $\mathcal{B}^0$ is implied by the observation that the above three-form is just the pullback, under the map $U$, of the volume form on target space (up to a normalization factor) \cite{Sp1}, \cite{term}, \cite{rho-book}, 
\be \label{pullback}
 n\,  {\rm vol}_{\mathbb{R}^3} = \frac{1}{2\pi^2} U^* \left(   {\rm vol}_{\mathbb{S}^3} \right) .
\ee
Indeed, it immediately follows that $\int n\,  {\rm vol}_{\mathbb{R}^3} = B (1/(2\pi^2)) \int {\rm vol}_{\mathbb{S}^3} =B$, where $2\pi^2$ is the volume of $\mathbb{S}^3$, and $B$ counts the number of times the target space is covered by the map $U$.  

The static energy functional is
\be
E_{\rm BPS} =\int   {\rm vol}_{\mathbb{R}^3} \rho = \int   {\rm vol}_{\mathbb{R}^3} \left( \lambda^2 \pi^4 n^2 + \mu^2 \mathcal{U} \right)
\ee
and allows for the usual completion of the square to derive the BPS bound and equation. For our purposes it is, however, simpler to use the conservation of the EMT, which, for static configurations in flat  space, is equivalent to the condition of constant pressure,
\be
p = \lambda^2 \pi^4 n^2 - \mu^2 \mathcal{U} \equiv P = \mbox{const} .
\ee
(the proper BPS case can be recovered in the limit $P \to 0$).
This condition is, in fact, a first integral of the static field equations, where $P$ is an integration constant.
It may be re-written like
\be \label{const-pressure-2}
 n = \pm \frac{\mu}{\lambda \pi^2} \sqrt{\mathcal{U} + (P/\mu^2)}
 \ee
 (in the sequel we will choose the $+$ sign). This equation (which generalizes the BPS equation to non-zero pressure) has a huge amount of symmetries, among them the volume-preserving diffeomorphisms (VPS) on physical space \cite{footnote2}. Inserting it into (\ref{pullback}) allows to re-express the invariant volume form as a pullback by itself,
 \be
   {\rm vol}_{\mathbb{R}^3} = \frac{\pi^2\lambda}{\mu} \frac{1}{2\pi^2} U^* \left(  \frac{ {\rm vol}_{\mathbb{S}^3} }{\sqrt{\mathcal{U} + (P/\mu^2)}}\right) .
\ee
This implies that for integrands which depend on $\vec x$ only via $U$, the resulting integrals can be written as target space integrals and give the {\em same} result for any static solution (equilibrium configuration) with the same pressure (integration constant) $P$. In particular, for the volume $V(P)$ and on-shell energy (i.e., using the constant pressure condition) we get
\be \label{V-P}
V(P) = \int {\rm vol}_{\mathbb{R}^3} = B \pi^2 \frac{\lambda }{\mu} \left< \frac{1}{\sqrt{{\cal U} + (P/\mu^2)}} \right>
\ee
\be \label{E-P}
E(P) = \int   {\rm vol}_{\mathbb{R}^3} \left( 2 \mu^2 \mathcal{U} +P \right) =
B \pi^2 \lambda \mu \left< \frac{2{\cal U} + (P/\mu^2)}{\sqrt{{\cal U} + (P/\mu^2)}} \right>
\ee
where the brackets define the target space average, $\left< f(U) \right> = (2\pi^2)^{-1} \int {\rm vol}_{\mathbb{S}^3} f(U)$. They obey the thermodynamical relation $P = -(dE/dV)$, as may be checked easily. For typical potentials like the pion mass potential $\mathcal{U}_\pi = 2h$ or the pion mass potential squared, these averages can be expressed in terms of generalized hypergeometric functions \cite{BPS-NS2}, \cite{rho-book}. 

The above results allow us to define average observables, corresponding to a MFT treatment of the BPS Skyrme model. In particular, the average energy density $\bar \rho$, baryon density $\bar n$ and baryon chemical potential $\bar{\boldsymbol{\mu}}$ for arbitrary static solutions are
\be \label{MF-eos}
\bar \rho (P) = \frac{E(P)}{V(P)} \, , \quad \bar n(P)  = \frac{B}{V(P)} \, , \quad \bar{\boldsymbol{\mu}}(P) = \frac{\bar \rho + P}{\bar n}
\ee
and lead to a {\em barotropic} perfect fluid, by construction.

{\em Remark:} 
Looking at relation (\ref{rho-p-n-rel}), it is tempting to define a local chemical potential density 
$\tilde{\boldsymbol{\mu}} = 2\pi^4 \lambda^2 n$. It should be emphasized, however, that  
$\tilde{\boldsymbol{\mu}}$ is {\em not} a chemical potential. In particular, it is not constant at equilibrium. In any case, by integrating Eq. (\ref{rho-p-n-rel}) we find a second, equivalent expression for $\bar{\boldsymbol{\mu}}$,
\be
\bar{\boldsymbol{\mu}} = B^{-1} \int {\rm vol}_{\mathbb{R}^3} \, n \tilde{\boldsymbol{\mu}} = 2\pi^4 \lambda^2 B^{-1} \int {\rm vol}_{\mathbb{R}^3} \, n^2 
= 2\pi^2 \lambda\mu \left< \sqrt{\mathcal{U} + (P/\mu^2)} \right> .
\ee

\subsection{Parameter values}
Before doing numerical calculations, we have to choose values for the parameters $\lambda$ and $\mu$ of the model. As explained in the introduction, we will do so by fitting to the properties of infinite nuclear matter. 
Concretely, we consider the two potentials $\mathcal{U}_\pi$ and $\mathcal{U}_\pi^2$, and a third potential which approaches the vacuum with a fourth power in $\xi$, like the potential $\mathcal{U}_\pi^2$, but is flat (constant) in the anti-vacuum hemisphere $1/2 \le h \le 1$, i.e., 
\be \label{flat-pot}
\mathcal{U}_{\rm flat} = \left \{ \begin{array}{ll}
\sin^4 \xi = 16 h^2 (1-h)^2 \; &, \; \xi \in [0,\frac{\pi}{2 }], \\
1 &, \; \xi \in [\frac{\pi}{2},\pi].
\end{array} \right.
\ee
The idea here is that the two potentials $\mathcal{U}_\pi$ and $\mathcal{U}_\pi^2$ are the simplest expressions in terms of the Skyrme field. 
They may be considered as the first two  terms in an expansion, which provide the leading contributions for small field (close to the vacuum).  In a more general model, probably both terms will contribute, but we chose the two extreme cases for simplicity.
Further, their approach to the vacuum is as required from physical considerations (the quadratic approach of $\mathcal{U}_\pi$ provides the pions with a mass, whereas the quartic approach of $\mathcal{U}_\pi^2$ induces a repulsive short-range interaction, alleviating the binding energy problem \cite{bjarke1}-\cite{bjarke3}). On the other hand, they are rather peaked at the anti-vacuum $h=1$, leading to large central baryon densities already in the case without gravitation. 
We, therefore, chose the flat potential ${\cal U}_{\rm flat}$ as an example which avoids these high central baryon densities.
For a flat potential like (\ref{flat-pot}), the baryon density is constant in the region of constant potential by construction, as an immediate consequence of the constant pressure equation (\ref{const-pressure-2}).  This behavior
probably provides a more realistic modeling of nuclear matter for high densities, and the potential (\ref{flat-pot}) is a concrete example thereof. In any case, the detailed determination of the correct Skyrme model potential will require a fitting to many more observables of nuclear physics.

The corresponding fit values are (for $\mathcal{U}_\pi$ and $\mathcal{U}_\pi^2$ they were calculated in \cite{BPS-NS2}, but for the slightly different values $n_0 = 0.153 \, {\rm fm}^{-3}$, $E_{\rm inm} = 923.3  \; {\rm MeV}$ for the parameters of infinite nuclear matter)
\bea \label{param-pot-2}
{\cal U}_\pi: &&   \lambda^2 = 26.88 \; {\rm MeV} \,{\rm fm}^3  ,  \; \;  \mu^2 = 88.26 {\rm fm}^{-3}\, {\rm fm}^{-3}  \\
 \label{param-pot-4}
{\cal U}_\pi^2: &&  \lambda^2 = 15.493 \; {\rm MeV} \,{\rm fm}^3  , \; \;   \mu^2 = 141.22 \; {\rm MeV}\, {\rm fm}^{-3} \\
{\cal U}_{\rm flat} : && \lambda^2 = 23.60 \; {\rm MeV} \,{\rm fm}^3  ,  \; \;  \mu^2 = 121.08 \; {\rm MeV}\, {\rm fm}^{-3} .
 \eea

\section{Numerical Neutron Star calculations}
For the numerical calculations, both in the MFT and in the exact BPS case, we use a shooting from the center. Details of the method and of the relevant boundary conditions in both cases can be found in \cite{BPS-NS2}.

\subsection{Full field theory calculations with the effective crust}
In the full field theory case, we calculate the NS masses and radii resulting from the effective crust description by a two-step procedure. In a first step, we determine the full BPS theory NS solutions by integrating the corresponding TOV equations from the center $r=0$ to the  BPS NS radius $R_{\rm BPS}$, defined as the radius where the full BPS theory pressure takes the value zero, 
$p_{\rm core} (R_{\rm BPS})\equiv  p_{\rm BPS} (R_{\rm BPS})=0$. In a second step, we determine the radius $r_{\rm cc}$ where the chemical potential from the full BPS theory takes the value $\boldsymbol{\mu}_{\rm cc} = 955\, \mbox{MeV}$, that is to say, $\boldsymbol{\mu}_{\rm core} (r_{\rm cc}) \equiv \boldsymbol{\mu}_{\rm BPS}(r_{\rm cc}) =\boldsymbol{\mu}_{\rm cc}$. For $r>r_{\rm cc}$, we then replace the BPS crust by the effective crust, using the procedure of \cite{ZFH} explained in Section III.
The resulting NS observables are the NS radius $R$ given by Eq. (\ref{crust-radius}) and the NS mass $M_{\rm NS} = M_{\rm core} + M_{\rm crust}$ where the crust contribution is given by (\ref{crust-mass}). Both observables are somewhat different from the full BPS theory values $R_{\rm BPS}$ and $M_{\rm BPS} \equiv m_{\rm BPS}(r=R_{\rm BPS})$.

More concretely, we use the expressions  (\ref{full-rho}) and (\ref{full-p}) for $\rho$ and $p$ in the TOV equations (\ref{Tov1}) and (\ref{Tov2a}) for the full field theory calculation. They constitute a system of two equations for the two unknown functions $h$ and $\bf{B}$ and, therefore, close by themselves. In this case, the baryon number $B$ is an input parameter on which $\rho$ and $p$ depend explicitly, see (\ref{full-rho}) and (\ref{full-p}). As a consequence, solutions do not exist for arbitrary initial values $p_0 \equiv p(r=0)$. Instead, for each $B$ the corresponding $p_0$ must be found such that 
the resulting formal solution is sufficiently regular at the BPS NS surface $R_{\rm BPS}$ and leads to non-singular metric functions there. Concretely, the regularity condition is that
at the BPS NS radius $R_{\rm BPS}$ (defined by $p_{\rm BPS} (R_{\rm BPS})=0$) of the pure BPS model (with BPS crust) the condition $p'(R_{\rm BPS})=0$ is satisfied, see \cite{BPS-NS1}, \cite{BPS-NS2}. Once this solution is found for a given $B$, the effective crust is then constructed as a second step.  

To determine this effective crust, we have to find the radius $r_{\rm cc}$ where $\boldsymbol{\mu}_{\rm BPS}(r_{\rm cc}) =\boldsymbol{\mu}_{\rm cc}$. 
 As explained in Section III, this faces the additional problem that for the exact BPS model, like for any non-barotropic fluid, only the differential relation (\ref{mu'}) for the chemical potential is available, which allows to determine the functional dependence of $\boldsymbol{\mu}_{\rm BPS} (r)$ but not its absolute value. This absolute value can be determined by the following observation. The classical soliton solutions of the BPS Skyrme model correspond to infinite nuclear matter, by assumption. But this implies that at the surface of the pure BPS Skyrme NS, where $p_{\rm{BPS}}(R_{\rm BPS})=0$, the chemical potential should be equal to the energy per baryon of infinite nuclear matter, $\boldsymbol{\mu}_{\rm{BPS}}(R_{\rm BPS})=\boldsymbol{\mu}_0 \equiv 922.9 \, \mbox{MeV}$. This determines the absolute value of $\boldsymbol{\mu}_{\rm BPS}$ and allows to find the radius $r_{\rm cc}$ where $\boldsymbol{\mu}_{\rm BPS}(r_{\rm cc})=955 \, \mbox{MeV}$. In a last step, for $r\ge r_{\rm cc}$ the BPS solution is replaced by the effective crust, as described in Section III.B. We show our results for the potentials considered in this paper in Fig. 1.  
   
In particular, for the potential $\mathcal{U}_\pi^2$ (central panel of Fig. 1) we find that the $M$-$R$ curves with and without the effective crust are almost the same for $M\ge 0.5 M_\odot$. The reason is that, for this potential with its quartic approach to the vacuum, the energy density in the "tail" or BPS crust region of the pure BPS NS (close to $R_{\rm BPS}$) is very small for the full BPS model. As a consequence, the "post-Newtonian" approximation implied by the procedure of \cite{ZFH} gives a result which is rather close to the full field theory result in this region.

\begin{figure}[h]
\begin{center}
\includegraphics[width=0.90\textwidth]{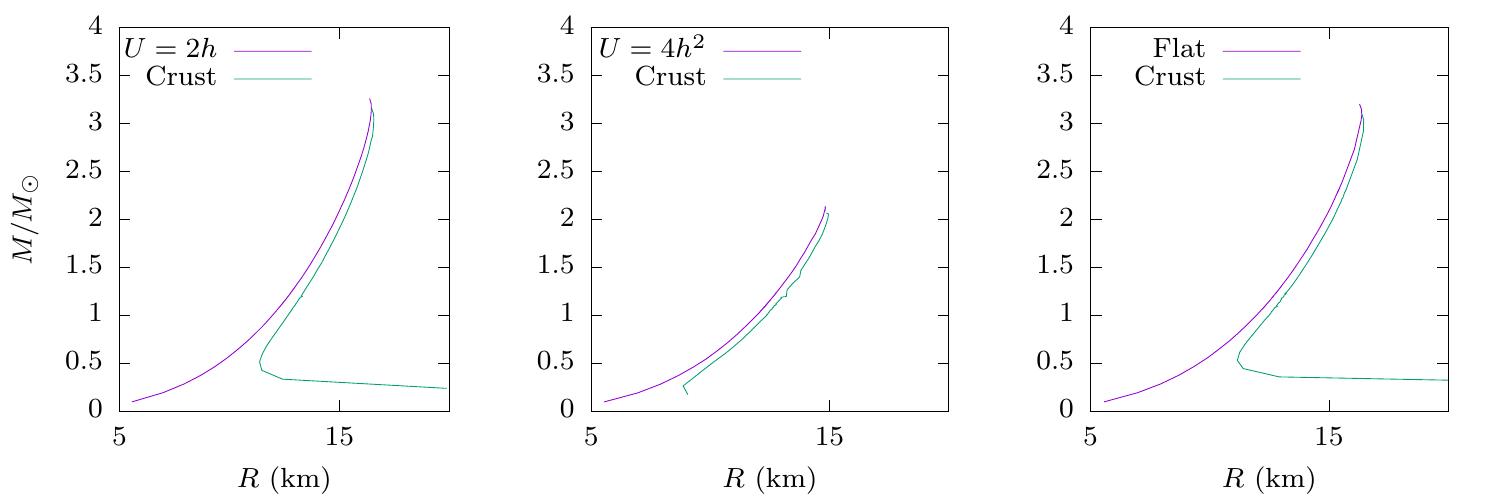}
\end{center}
\caption{Mass-radius relation for NS of the BPS Skyrme model in full field theory, for the three potentials $\mathcal{U}_\pi = 2h$, $\mathcal{U}_\pi^2 = 4h^2$ and ${\mathcal U}_{\rm flat}$. For comparison, we show both the case with (green line) and without (violet line) the effective crust.}
\end{figure}
\subsection{Mean field theory calculations joining the BCPM EoS of Ref. \cite{vinas}}
Even in the MFT case, one might want to try the simple effective crust construction of \cite{ZFH} as a first, rough approximation. This construction is, however, no longer consistent in the MFT case. Indeed, the MFT BPS Skyrme NS ends at a density $n(R)=n_0$, therefore the putative "crust" must be joined at some $n_* > n_0$. If this $n_*$ is chosen very close to $n_0$, then the resulting "crust" thickness is unrealistically small. If, on the other hand, $n_*$ is chosen sufficiently large to provide a reasonable crust thickness, then the corresponding crust mass is no longer small, and the self-consistency of the whole construction breaks down.

In the MFT case, therefore, we join the EoS resulting from the MFT BPS Skyrme model for $n>n_*$ to the BCPM EoS of \cite{vinas} for $n<n_*$, for some $n_* > n_0$, as explained in the introduction. In other words, we assume that the EoS of \cite{vinas} is valid both for the crust $n<n_{\rm cc}$ and for the outer core $n_{\rm cc} < n <n_*$, whereas we use the MFT BPS Skyrme EoS for the inner core $n>n_*$. For simplicity, we will use the corresponding pressure value $p_* = p(n_*)$ to fix the transition point. In addition, in some cases we will assume a smooth interpolation between the two EoS instead of a sudden transition at a fixed $n_*$ (or $p_*$), concretely
\be \label{interpol}
\rho_{\rm tot} (p) = (1-\alpha (p)) \rho_{\rm BCPM} (p) + \alpha (p) \rho_{\rm BPS} (p)
\ee
where $\rho_{\rm BPS}(p)$ is given by the expression $\bar \rho (p)$ of Eq. (\ref{MF-eos}),  
\be
\alpha (p) = \frac{b^2p^2}{1+b^2 p^2}
\ee
is the interpolation function, and $b$ is a parameter which determines the transition region (located close to $p_* \sim (1/b)$). We show our numerical results in Figs. 2 and 3, which are discussed in some more detail in the next section.

 The BCPM EoS for the (liquid) core in \cite{vinas} is derived using some advanced standard methods of many-particle nuclear physics based on the Brueckner-Hartree-Fock (BHF) approach (plus the BCPM density functional for the crust), where nucleons are treated as quantum mechanical point particles. For the outer core  $n_{\rm cc} < n <n_*$, nuclear matter is rather well understood, and the EoS of \cite{vinas} can be expected to provide a precise description of NS matter. For densities which are significantly higher than $n_0$, on the other hand, the EoS of \cite{vinas} - like all other proposals - must be considered an {\em extrapolation}. One basic assumption of this extrapolation is the adequacy of a point-particle description of nucleons even at high density. Skyrme models, on the other hand, are paradigmatic examples of EFT where nucleons are extended objects (topological solitons), and their extended nature is the underlying cause for the rather high stiffness of the resulting nuclear matter at large densities. A Skyrme model treatment of nuclear matter, therefore, allows us to confront these different assumptions and to scrutinize their consequences for nuclear matter at high density and for the resulting NS properties. We remark that one common assumption of the EoS of \cite{vinas} and of the Skyrme model is the absence of additional degrees of freedom (hyperons, quarks, etc.) in the interior of NS (this was one reason to choose the EoS of \cite{vinas} for our comparison, in addition to its rather up-to-date character).

\begin{figure}[h]
\begin{center}
\includegraphics[width=0.90\textwidth]{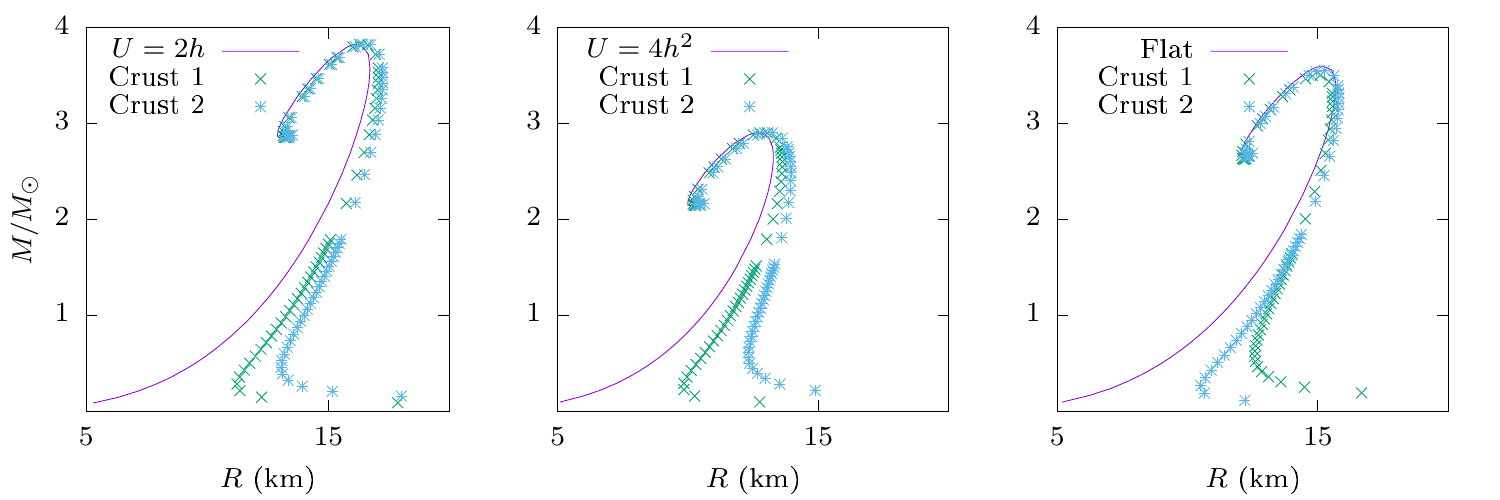}
\end{center}
\caption{Mass-radius relation for NS of the BPS Skyrme model in the MFT limit, for the three potentials $\mathcal{U}_\pi = 2h$, $\mathcal{U}_\pi^2 = 4h^2$ and ${\mathcal U}_{\rm flat}$. For the potentials $2h$ and $4h^2$, we choose a sudden transition between the BCPM EoS of Ref. \cite{vinas} and the BPS EoS.
Concretely, for  $\mathcal{U}_\pi = 2h$ at $p_* = 0.1 ({\rm  MeV}/{\rm fm}^3)$ (Crust 1) and
at $p_* = 2.5 ({\rm  MeV}/{\rm fm}^3)$ (Crust 2), and for $\mathcal{U}_\pi^2 = 4h^2$ at 
$p_* = 0.1 ({\rm  MeV}/{\rm fm}^3)$ (Crust 1) and at $p_* = 8.0 ({\rm  MeV}/{\rm fm}^3)$ (Crust 2).
For the potential ${\mathcal U}_{\rm flat}$ we choose the smooth interpolation (\ref{interpol}) with the two parameter values $b=0.1 \, ({\rm fm}^3/{\rm MeV})$ (Crust 1) and $b=10 \, ({\rm fm}^3/{\rm MeV})$ (Crust 2). 
For comparison, we also show the pure BPS case without the crust. For reasons of numerical convenience, also the unstable branch (where $M$ diminishes with increasing central pressure) is shown.}
\end{figure}

\begin{figure}[h]
\begin{center}
\includegraphics[width=0.90\textwidth]{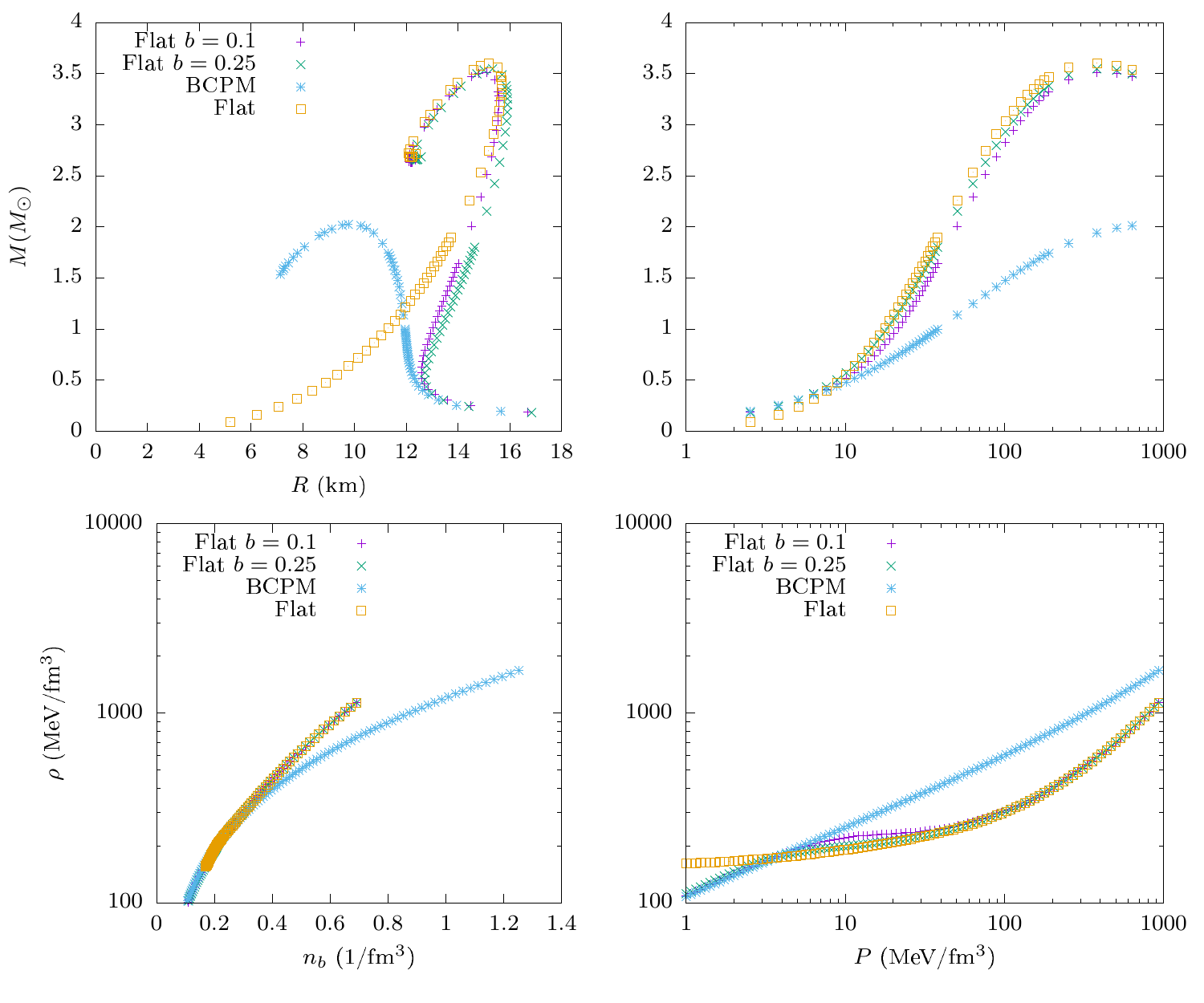}
\end{center}
\caption{Smooth interpolation (\ref{interpol}) for the potential ${\mathcal U}_{\rm flat}$, for the three values of the interpolation parameter $b=0$ (pure BPS case, "Flat"), $b= 0.1\, ({\rm fm}^3/{\rm MeV})$ and $b= 0.25 \, ({\rm fm}^3/{\rm MeV})$. In addition to the $M(R)$ curve (left upper panel) we also show the NS mass as a function of the central pressure (right upper panel) and the two EoS $\rho (n_b)$ (left lower panel) and $\rho (p)$ (right lower panel). $n_b$ is the baryon number density. For comparison, we also show the case of the BCPM EoS of Ref. \cite{vinas}. }
\end{figure}

 \section{Conclusions and Discussion}
 It was the main purpose of the present paper to 
 discuss possible ways to add a crust to
 skyrmionic NS and, in particular, to NS described by the BPS Skyrme model. In the BPS Skyrme model, both an exact and a MFT calculation of NS are possible and must be distinguished. For the exact case, we find that the method of \cite{ZFH} can be applied without problems and leads to the following typical behavior. The crust contribution to the total NS radius is significant only for rather low-mass NS (below one solar mass). This is related to the high stiffness of nuclear matter described by the BPS Skyrme model. At high density, the BPS Skyrme model reaches the maximally stiff EoS $\rho \sim p+\mbox{const}.$, and this limit is approached rather soon. The high stiffness also explains two further results which can be appreciated in fig. 1, namely the rather high maximum NS masses and the fact that the radius increases with the NS mass in the region between one and two solar masses. 
Concretely, we chose the BPS Skyrme model for the three potentials $\mathcal{U}_\pi$, $\mathcal{U}_\pi^2$ and ${\cal U}_{\rm flat}$, for the reasons explained in section IV.B. It turns out that the resulting bulk NS properties are qualitatively similar for the three cases. NS with 1.5 solar masses have radii between 12 and 14 km, and NS with 2 solar masses have radii between 13 and 15 km. 

In the case of a MFT description of BPS Skyrme NS, a crust cannot be attached directly to the BPS Skyrme NS using the method of \cite{ZFH}. Instead, the EoS of the BPS Skyrme model must be complemented with a standard nuclear physics EoS at low densities. Concretely, we chose the BCPM EoS of \cite{vinas}. Even so, bulk properties like
the $M(R)$ curves are not very different between the exact and MFT cases. We want to emphasize, however, that coordinate-dependent observables like densities differ considerably between MFT and exact calculations \cite{BPS-NS2}, particularly for the potentials $\mathcal{U}_\pi$, $\mathcal{U}_\pi^2$ considered in this letter, see footnote \cite{footnote3} below. 
 
 A further consequence of the high stiffness of the BPS Skyrme model is that the baryon densities in the interior of a NS remain relatively low. Indeed, as can be calculated easily, for the models $\mathcal{U}_\pi$ and ${\cal U}_{\rm flat}$ even in the center $r=0$ the baryon density never increases by more than a factor of about three
 in comparison with the case without gravity \cite{footnote3}. For the potential ${\cal U}_{\rm flat}$  this can be easily inferred from Fig. 3. The exception is the model $\mathcal{U}_\pi^2$ where the baryon density at the center grows to about 4.5 times the saturation density $n_0$ for the maximum mass NS \cite{BPS-NS1,BPS-NS2}. This is related to the fact that $\mathcal{U}_\pi^2$ is rather peaked around the anti-vacuum, and may imply that potentials which are flatter there are more realistic. We introduced the potential ${\cal U}_{\rm flat}$ (which approaches the vacuum like $\mathcal{U}_\pi^2$ but is flat in the large-field region) precisely for this reason.
 
Indeed, baryon densities which are not excessively large in the interior of a skyrmionic NS are important for the self-consistency of the Skyrme model approach to NS. The Skyrme model only describes standard nuclear matter, by construction. There are no contributions from exotic hadrons, and a dissolution of baryonic matter into quarks does not occur. But this assumption would be rather unlikely in an environment of extremely high baryon density. If, on the other hand, the baryon density never exceeds several times the density of normal (non-gravitating)  nuclear matter, then this assumption is much more plausible. 
Within the generalized Skyrme model (\ref{gen-sk}), the sextic term $\mathcal{L}_6$ is responsible for this behavior \cite{Sk-eos}, which underlines its importance. It describes a strong repulsion acting on compressed nuclear matter as a result of the strong interaction. In RMF models of nuclear matter, this repulsive force is induced by the omega meson. In other words, a stiff EoS, rather low baryon densities and a description entirely in terms of standard nuclear matter are characteristic features of a skyrmionic approach to NS, which distinguishes it from many other approaches. The observed high-mass NS with $M> 2M_\odot$, which require rather stiff nuclear matter, are a strong argument in favor of this approach, all the more so because the inclusion of additional degrees of freedom tends to soften the EoS.
The same line of reasoning also explains the preliminary finding that isospin contributions to the skyrmionic NS may be ignored in the NS core. It is equivalent to the statement that for compressed nuclear matter the repulsive force induced by the strong interactions is much more important than the degeneracy pressure. Of course, these arguments do not apply to the NS crust, but the effective crust description of \cite{ZFH} was introduced precisely with the aim of avoiding a detailed description of the crust, which does not seem to be very important for NS bulk properties.

The BPS submodel of the generalized Skyrme models (\ref{gen-sk}) is singled out both because of its simplicity and because, owing to its stiffness, it provides the leading contribution for the inner core. As a next step towards a more complete description of NS, the full model (\ref{gen-sk}) should be considered. In this case, an ansatz leading to a spherically symmetric energy density is no longer available (except for baryon number $B=1$).  A full field theoretic treatment is, therefore, beyond current possibilities. As to a MFT approach, we already know two limiting cases, namely the perfect fluid of the BPS submodel and the crystal provided by the $\mathcal{L}_2 + \mathcal{L}_4$ submodel. It follows that the full generalized model (\ref{gen-sk}) will show a rather complex pattern also in MFT, possibly with several topological phase transitions, depending both on parameter values and on baryon density. Taking into account the scaling behavior of the different terms, however, it is plausible to assume that the inner core will continue to be described by a fluid, whereas the outer core will show some type of crystalline structure. Owing to its slightly softer overall EoS, the full model will also lead to smaller maximum NS masses and to slightly smaller NS radii for a given NS mass, which might be desired features. In particular, recent observations of gravitational waves emitted from inspiraling NS binaries seem to imply an upper bound on the maximum NS mass of about $M_{\rm max} \sim 2.3 M_\odot$ \cite{Shibata} and a value of $R\sim 11\,$km for a $1.4 M_\odot$ NS \cite{capano}. The pure $\mathcal{L}_2 + \mathcal{L}_4$ Skyrme crystal leads to a maximum NS mass of about 1.9 $M_\odot$ and to a maximum radius of about 11 km \cite{piette3}, \cite{naya}, therefore an appropriate combination of the standard and BPS Skyrme models should be able to naturally accommodate these most recent constraints.

\section*{Acknowledgements}
The authors acknowledge financial support from the Ministry of Education, Culture, and Sports, Spain (Grant No. FPA2017-83814-P), the Xunta de Galicia (Grant No. INCITE09.296.035PR and Conselleria de Educacion), the Spanish Consolider-Ingenio 2010 Programme CPAN (CSD2007-00042), Maria de Maetzu Unit of Excellence MDM-2016-0692, and FEDER. Further, we would like to thank all those who made it possible for us to work safely at home in these times of Covid19.


\end{document}